\documentclass[12pt,dvips]{article}
\usepackage[dvips]{graphicx}
\usepackage{graphicx}

\graphicspath{{plots/}}
\DeclareGraphicsExtensions{.eps.gz,.eps,.ps,.ps.gz}

\parskip=\itemsep               %?
\newcommand{\lwig}{\mbox{\,\raisebox{.3ex}
    {$<$}$\!\!\!\!\!$\raisebox{-.9ex}{$\sim$}\,}}
\newcommand{\gwig}{\mbox{\,\raisebox{.3ex}
    {$>$}$\!\!\!\!\!$\raisebox{-.9ex}{$\sim$}}\,}
\newcommand{\lambdabar}{{\hbox{$\lambda_e$\kern-1.9ex\raise+0.45ex\hbox{--}
\kern+0.2ex}}}

\date{\empty}

\title{{\normalsize\rightline{DESY 01-024}\rightline{hep-ph/0103185}}
\vskip 1cm 
\bf Pair Production from Vacuum at the Focus of an X-Ray Free Electron Laser  
       \vspace{21mm}} 
\author{A. Ringwald\\[4mm] 
Deutsches Elektronen-Synchrotron DESY, Hamburg, Germany}

\begin{document}
\begin{titlepage} 
  \maketitle
% declarations for front matter
\vspace{3cm}
\begin{abstract}
There are definite plans for the construction of X-ray 
free electron lasers (FEL), both at DESY, where the so-called 
XFEL is part of the design of the electron-positron linear 
collider TESLA, as well as at SLAC, where the so-called Linac Coherent Light 
Source (LCLS) has been proposed. Such an X-ray laser would allow for 
high-field science applications: One could make use of not only the high 
energy and transverse coherence of the X-ray beam, but also of 
the possibility of focusing it to a spot with a small radius, 
hopefully in the range of the laser wavelength. Along this route one  
obtains very large electric fields, much larger than those obtainable with any 
optical laser of the same power. In this letter we discuss 
the possibility of obtaining an electric field so high that electron-positron 
pairs are spontaneously produced in vacuum (Schwinger pair production). We 
find that if X-ray optics can be improved to approach the 
diffraction limit of focusing, and if the power of the planned X-ray FELs 
can be increased to the terawatt region, then there is ample room for 
an investigation of the Schwinger pair production mechanism.      
\end{abstract}


\thispagestyle{empty}
\end{titlepage}
\newpage \setcounter{page}{2}

%\section{Introduction}
{\em 1.}
Spontaneous particle creation from vacuum induced by an external field, first 
put forth to examine the production of 
electron-positron ($e^+e^-$) pairs in a static, spatially uniform electric 
field~\cite{Sauter:1931,Heisenberg:1936qt,Schwinger:1951nm} and often 
referred to as the Schwinger mechanism, ranks among the 
most 
intriguing nonlinear phenomena in quantum field theory. Its consideration is
theoretically important, since it requires one to go beyond 
perturbation theory, and its experimental observation would 
verify the validity of the theory in the domain of strong fields.  
Moreover, this mechanism has been applied to many problems in 
contemporary physics, ranging from black hole quantum 
evaporation~\cite{Hawking:1975sw,Damour:1976jd,Gibbons:1978pt,Gavrilov:1996pz}
to particle production in hadronic 
collisions~\cite{Casher:1979wy,Andersson:1983ia,Biro:1984cf} 
and in the early 
universe~\cite{Parker:1969au,Birrell:1982}, to 
mention only a few. One may consult the 
monographs~\cite{Greiner:1985,Grib:1988,Fradkin:1991} for a review of further 
applications, concrete calculations and a detailed bibliography.

It is known since a long time that in the background of a static, spatially 
uniform electric field the vacuum in quantum electrodynamics (QED) is unstable
and, in principle, sparks with spontaneous emission of $e^+e^-$ 
pairs~\cite{Sauter:1931,Heisenberg:1936qt,Schwinger:1951nm}.
However, a sizeable rate for spontaneous pair production requires 
extraordinary strong electric field strengths $\mathcal E$ of order or
above the critical value
\begin{equation}
{\mathcal E}_c \equiv \frac{m_e\, c^2}{e\, \lambdabar} 
= \frac{m_e^2\, c^3}{e\, \hbar}
\simeq 1.3\times 10^{18}\ {\rm V/m}\,. 
\label{schwinger-crit}
\end{equation}
Otherwise, for $\mathcal E\ll \mathcal E_c$, the work of the 
field on a unit charge $e$ over the Compton wavelength of the electron 
$\lambdabar =\hbar /(m_e c)$ is much smaller than the rest energy 
$2\,m_e c^2$ of the produced $e^+e^-$ pair, the
process can occur only via quantum tunneling, and its rate is  
exponentially suppressed, $\propto \exp (- \pi \mathcal E_c/\mathcal E)$.

Unfortunately, it seems inconceivable to produce macroscopic 
static fields 
with electric field strengths of the order of the Schwinger critical 
field~(\ref{schwinger-crit}) in the laboratory. In view of this difficulty,    
in the early 1970's the question was raised\footnote{At about the same time, 
the thorough investigation of the question started whether the necessary 
superstrong fields around ${\mathcal E}_c$ can be generated microscopically and
transiently in the Coulomb field of colliding heavy ions with 
$Z_1+Z_2 > Z_c\approx 170$~\cite{Zeldovich:1972,Muller:1972}. At the present 
time, clear experimental signals for spontaneous positron creation in heavy 
ion collisions are still missing and could
only be expected from collisions with a prolonged lifetime (for a recent
status report of this issue, see Ref.~\cite{Greiner:1998}).}  
whether intense optical lasers could be employed to 
study the Schwinger mechanism~\cite{Bunkin:1970,Brezin:1970}. Yet, it was 
found that all available and conceivable optical lasers did not have enough 
power density to allow for a sizeable pair creation rate~\cite{Bunkin:1970,
Brezin:1970,Popov:1971,Popov:1972,Troup:1972,Popov:1972b,Popov:1973,
Narozhnyi:1974,Popov:1974,Mostepanenko:1974,Popov:1974b,Marinov:1977gq}.

Meanwhile, there are definite plans for the construction of X-ray 
free electron lasers (FEL), both at DESY, where the so-called 
XFEL is part of the design of the $e^+e^-$ linear 
collider TESLA~\cite{Brinkmann:1997nb,Materlik:1999uv,tesla-tdr}, as well as 
at SLAC, where the so-called Linac Coherent Light Source (LCLS) has been 
proposed~\cite{Arthur:1998yq,Lindau:1999uw}. 
It has been pointed out by several 
authors~\cite{Melissinos:1998qn,Chen:1998,Chen:1999kp,Tajima:2000} that 
an X-ray laser would allow for high-field science applications: 
One could make use of not only the high energy and transverse coherence of 
the X-ray beam, but also of 
the possibility of focusing it to a spot with a small radius $\sigma$, 
hopefully in the range of the laser wavelength, 
$\sigma\,\gwig\, \lambda\simeq \mathcal{O}(0.1)$~nm. In this way one might 
obtain very large electric fields, $\mathcal E\propto 1/\sigma\sim 1/\lambda$,
much larger than those obtainable with any optical laser of the same power. 

Electron-positron pair production at the focus of an X-ray FEL has been 
discussed in Ref.~\cite{Melissinos:1998qn}\footnote{In 
Ref.~\cite{Melissinos:1998qn} the production of positrons  
in the collision of 46.6 GeV/c electrons with terawatt optical laser 
pulses, observed by the SLAC experiment
E-144~\cite{Burke:1997ew}, was discussed. Whereas in Ref.~\cite{Burke:1997ew}
the data were interpreted in terms of multiphoton light-by-light scattering,
an alternative explanation in terms of the Schwinger mechanism was offered
in Ref.~\cite{Melissinos:1998qn} (see also Ref.~\cite{Bamber:1999zt}).} 
and an estimate of the corresponding rate has been presented in 
Ref.~\cite{Chen:1998}. 
It is the purpose of this letter to strengthen these considerations and to 
present a state of the art evaluation of the prospects to observe the 
Schwinger mechanism at future X-ray laser facilities. 
In particular, we determine critical laser parameters, like the 
laser power and the focus spot size, which should be aimed at to 
get an observable effect. To this end, we make heavily use of the rather well 
forgotten work on the Schwinger mechanism in alternating electric fields by 
Russian groups~\cite{Popov:1971,Popov:1972,Popov:1972b,Popov:1973,
Narozhnyi:1974,Popov:1974,Mostepanenko:1974,Popov:1974b,Marinov:1977gq}.   

%\section{The Model}
{\em 2.}
We start with a discussion of a number of simplifying approximations 
concerning the electromagnetic field of the laser radiation.
We elaborate on a model which retains the
main features of the general case but nevertheless allows to
obtain final expressions for the pair production rate in closed form. This 
should be sufficient for  
an order-of-magnitude estimate of the critical parameters.

It is well known that no pairs are
produced in the background of a light-like static, spatially uniform 
electromagnetic field~\cite{Schwinger:1951nm}, 
characterized invariantly by\footnote{Unless otherwise stated, we use the 
rationalized MKSA unit system throughout. For the numerical values of the
physical constants we take the ones given in Ref.~\cite{Groom:2000in}.}
\begin{eqnarray}
\label{F}
\mathcal F &\equiv & \frac{1}{4}\,F_{\mu\nu}F^{\mu\nu}\equiv 
-\frac{1}{2}\,(\mathbf E^2 -c^2\mathbf B^2)=0\,,
\\[1ex] 
\label{G}
\mathcal G &\equiv & \frac{1}{4}\,F_{\mu\nu}\tilde F^{\mu\nu}\equiv 
c\,\mathbf E\cdot\mathbf B =0\,,
\end{eqnarray}
where $F^{\mu\nu}$ is the electromagnetic field strength tensor and 
$\tilde F^{\mu\nu} =(1/2)\,\epsilon^{\mu\nu\alpha\beta}F_{\alpha\beta}$ its 
dual. It has been argued that 
fields produced in (optical) focusing of laser beams are very close to
such a light-like electromagnetic field, leading to an essential suppression 
of pair creation\footnote{Yet, in a focused wave there are regions near 
the focus where 
$\mathcal F<0$ and pair production is 
possible~\cite{Bunkin:1970,Melissinos:1998qn}.}~\cite{Troup:1972}. 
For other fields, $\mathcal F$ and $\mathcal G$ do not vanish, and pair
production becomes possible, unless $\mathcal G =0$, $\mathcal F>0$, 
corresponding to a pure magnetic field in an appropriate coordinate 
system~\cite{Schwinger:1951nm}. 
In particular, one expects pair creation in the background of 
a spatially uniform 
electric field oscillating with a frequency $\omega$, say
\begin{equation}
{\mathbf E}(t) =(0,0,{\mathcal E}\cos (\omega t))\,,
\hspace{6ex}
{\mathbf B}(t) =(0,0,0)\,,
\label{electric-type}
\end{equation}
which has $\mathcal G =0$, $\mathcal F<0$. As emphasized in 
Refs.~\cite{Popov:1972b,Popov:1973,Mostepanenko:1974,Marinov:1977gq,
Chen:1998}, such a field may be created in an antinode of the standing wave 
produced by a superposition of two coherent laser beams with wavelength 
\begin{equation}
\lambda = \frac{2\pi c}{\omega}\,,
\end{equation}
and, indeed, it may be considered as 
spatially uniform at distances much less than the wavelength.

Thus, for definiteness, we assume that every X-ray laser pulse is 
split into two equal parts and recombined to form a standing wave with
locations where the electromagnetic field has the 
form~(\ref{electric-type}) and 
where the peak electric field is given by (1 TW = $10^{12}$ W)
\begin{equation}
\label{peak-electric-field}
\mathcal E =\sqrt{
\mu_0\,c\,
\frac{P}{\pi \sigma^2} }
\ \simeq\ 1.1\times 10^{17}\ 
\left( \frac{P}{1\ {\rm TW}}\right)^{1/2}\,
\left( \frac{0.1\ {\rm nm}}{\sigma}\right)
\ \frac{\rm V}{\rm m}
\,,
\end{equation}
in terms of the laser power $P$, the focus spot radius $\sigma$ and the 
permeability of free space $\mu_0=4\pi\times 10^{-7}$ N A$^{-2}$. 
Furthermore, we assume that the peak electric field $\mathcal E$ is much 
smaller than the Schwinger critical field~(\ref{schwinger-crit})
and the energy of the laser photons is much smaller than the rest energy of 
the electron, 
\begin{equation}
{\mathcal E}\ll {\mathcal E}_c =  \frac{m_e^2\, c^3}{e\, \hbar}
\,,\hspace{10ex}
\hbar\omega\ll m_e c^2\,;
\label{conditions}
\end{equation}
conditions which are well satisfied at realistic optical as well as 
X-ray lasers (c.\,f. Table~\ref{parameters}). 
%%%%%%%%%%%%%%%%%%%%%%%%%%%Table%%%%%%%%%%%%%%%%%%%%%%%%%%%%%%%%%%%%%%
\begin{table} 
{\footnotesize
\begin{center}
\begin{tabular}{|ll|c||c|c|c|}\hline 
\multicolumn{6}{|c|}{{\bf Laser Parameters}}\\\hline\hline
 & & {\bf Optical}~\cite{Perry:1994}  & \multicolumn{3}{c|}{\bf X-ray FEL}
\\\hline
 & & Focus: & Design~\cite{tesla-tdr} & Focus: & 
Focus: \\  
   & & Diffraction limit &  & Available~\cite{Graeff:priv} & 
Goal~\cite{Chen:1998}  
\\\hline 
Wavelength& $\lambda$ & 1 $\mu$m & 0.4 nm & 0.4 nm & 0.15 nm \\%\hline
Photon energy & $\hbar\,\omega = \frac{hc}{\lambda}$ & 1.2 eV & 3.1 keV 
& 3.1 keV 
& 8.3 keV\\%\hline 
Peak power& $P$ & 1 PW & 110 GW  & 1.1 GW &  5 TW                \\%\hline
Spot radius (rms)& $\sigma$ & 1 $\mu$m & 26 $\mu$m & 21 nm  & 0.15 nm\\%\hline 
Coherent spike length (rms) & $\triangle t $ & 500 fs $\div$ 20 ps
& 0.04 fs%~\cite{Graeff:priv} 
& 0.04 fs%~\cite{Graeff:priv}  
& 0.08 ps               \\\hline\hline
\multicolumn{6}{|c|}{{\bf Derived Quantities}}\\\hline\hline
 & & & & & \\[0.5ex]
Peak power density&$S=\frac{P}{\pi \sigma^2 }$& $3\times 10^{26}$ 
$\frac{\rm W}{{\rm m}^2}$
& $5\times 10^{19}$ $\frac{\rm W}{{\rm m}^2}$ & 
$8\times 10^{23}$ $\frac{\rm W}{{\rm m}^2}$ & $7\times 10^{31}$ 
$\frac{\rm W}{{\rm m}^2}$
\\[1ex]%\hline
Peak electric field&$\mathcal E =\sqrt{\mu_0\,c\, S}$
& $4\times 10^{14}$ $\frac{\rm V}{\rm m}$ & 
$1\times 10^{11}$ $\frac{\rm V}{\rm m}$ &
$2\times 10^{13}$ $\frac{\rm V}{\rm m}$ & 
$2\times 10^{17}$ $\frac{\rm V}{\rm m}$
\\[1ex]%\hline
Peak electric field/critical field
 & $\mathcal E/{\mathcal E}_c$  &$3\times 10^{-4}$ & $1\times 10^{-7}$   
& $1\times 10^{-5}$  & 0.1     
\\%\hline
Photon energy/$e$ rest energy  &$\frac{\hbar\omega}{m_e c^2}$ & 
 $2\times 10^{-6}$ & $0.006$ &
$0.006$ & $0.02$ \\%\hline
Adiabaticity parameter & $\gamma = 
\frac{\hbar \omega}{e\, \mathcal E \lambdabar}$ 
& $9\times 10^{-3}$ & $6\times 10^{4}$ &
$5\times 10^{2}$ & 0.1 \\%\hline 
\hline
\end{tabular}
\caption[dum]{%\footnotesize 
Laser parameters and derived quantities relevant for 
estimates of the rate of spontaneous $e^+e^-$ pair production. 
The column labeled ``Optical'' lists parameters which are typical for
a petawatt-class (1 PW = $10^{15}$ W) optical laser~\cite{Perry:1994}, 
focused to the diffraction limit, 
$\sigma = \lambda$. The column labeled ``Design'' displays design 
parameters of the planned XFEL at DESY 
(``SASE-5'' in Ref.~\cite{tesla-tdr}). Similar values apply for 
LCLS~\cite{Arthur:1998yq,Lindau:1999uw}.
The column labeled ``Focus: Available'' shows typical values which can
be achieved with present day methods of X-ray 
focusing~\cite{Graeff:priv,Hastings:2000}: It assumes that the XFEL 
X-ray beam can be focused to a rms spot radius of $\sigma \simeq 21$ nm with 
an energy extraction efficiency of 1 \%~\cite{Graeff:priv}. The column 
labeled ``Focus: Goal'' 
shows parameters which are theoretically possible by increasing the energy 
extraction of LCLS (by the tapered undulator technique) and by a yet 
unspecified method of diffraction-limited focusing of X-rays~\cite{Chen:1998}.
\label{parameters}} 
\end{center}
}
\end{table}
%%%%%%%%%%%%%%%%%%%%%%%%%%%%%%%%%%%%%%%%%%%%%%%%%%%%%%%%%%%%%%%%%%%%%%

%\section{Rate Estimates: Theory}
{\em 3.}
Under these conditions, it is possible to compute the rate of 
$e^+e^-$ pair production in a semiclassical manner, using 
generalized WKB~\cite{Brezin:1970} or 
imaginary-time~\cite{Popov:1971,Popov:1972,Popov:1974,Popov:1974b} methods
(see also Ref.~\cite{Dunne:1998ni} and references cited therein).  
Let us summarize the basic results of the corresponding studies.
  
In Ref.~\cite{Brezin:1970}, the probability that an $e^+e^-$ pair is 
produced per unit time and unit volume,    
\begin{eqnarray}
w = 
\frac{{\rm d}\,n_{e^+e^-}}{{\rm d}^3x\,{\rm d}t}    
\,,
\end{eqnarray}
was estimated as
\begin{eqnarray}
\label{brezin-rate}
w_{\rm B\,I} =  
\frac{c}{4\,\pi^3 \lambdabar^4} \left( 
\frac{\mathcal E}{{\mathcal E}_c}\right)^2\,
\frac{\pi}{g(\gamma ) +\frac{1}{2}\,\gamma\,g^\prime (\gamma )}\,
\exp \left[ -\pi\, \frac{{\mathcal E}_c}{\mathcal E}\, g(\gamma ) \right]\,,
\end{eqnarray}
with~\cite{Brezin:1970,Popov:1971}
\begin{eqnarray}
g (\gamma ) \equiv  
\frac{4}{\pi}\int\limits_{0}^{1} {\rm d}u\,
\left[ 
\frac{1-u^2}{1+\gamma^2 u^2}
\right]^{1/2}
&=&
\frac{4}{\pi} \frac{\sqrt{1+\gamma^2}}{\gamma^2}
\left[ 
\mbox{\boldmath $K$} \left( \frac{\gamma}{\sqrt{1+\gamma^2}}\right)
- 
\mbox{\boldmath $E$} \left( \frac{\gamma}{\sqrt{1+\gamma^2}}\right)
\right]
\\[2ex]
&=&
\left\{
\begin{array}{lcr}
1-\frac{1}{8}\gamma^2+\mathcal O(\gamma^4)&:& \gamma\ll 1\,,\\[1.5ex]
\frac{4}{\pi\gamma}\ln \left( \frac{4\gamma}{\rm e}\right) 
+\mathcal O(1/\gamma^3) 
&:& \gamma\gg 1\,.
\end{array}
\right.
\label{g-asy}
\end{eqnarray} 
Here, {\boldmath $K$ and $E$} are the complete elliptic integrals of
the first and second kind, respectively. 

The laser frequency $\omega$ enters in the semiclassical 
formula~(\ref{brezin-rate}) 
only through the ratio $\gamma$ of the energy of the laser photons over 
the work of the field on a unit charge $e$ over the Compton wavelength of the 
electron,
\begin{equation}
\label{gamma}
\gamma \equiv \frac{\hbar\,\omega}{e\, {\mathcal E} \lambdabar }=
\frac{\hbar\,\omega}{m_e c^2}\,\frac{{\mathcal E}_c}{\mathcal E} =
\frac{m_e c\,\omega}{e\,\mathcal E}
\equiv \frac{{\mathcal E}_\omega}{\mathcal E}\,,
\end{equation}
which can take on arbitrary values despite of the 
conditions~(\ref{conditions}). 
This ratio plays the r\^ole of an adiabaticity parameter, 
as is easily inferred from the asymptotic behaviour~(\ref{g-asy}) of the 
function $g(\gamma )$, which enters in the principal, exponential factor in
the pair production probability~(\ref{brezin-rate}).   
Indeed, as long as $\gamma\ll 1$, 
i.\,e. in the high-field, low-frequency limit, formula~(\ref{brezin-rate})
agrees with the nonperturbative result from Ref.~\cite{Schwinger:1951nm} 
for a static, spatially uniform field,
\begin{equation}
\label{schwinger-rate}
w_{\rm S}    =   
\frac{c}{4\,\pi^3 \lambdabar^4} \left( 
\frac{\mathcal E}{{\mathcal E}_c}\right)^2\,
\sum_{\ell =1}^\infty \frac{1}{\ell^2}\,
\exp \left[ -\ell\,\pi\, 
\frac{{\mathcal E}_c}{\mathcal E} \right]\ 
\stackrel{{\mathcal E}\ll {\mathcal E}_c}{\simeq}\  
\frac{c}{4\,\pi^3 \lambdabar^4} \left( 
\frac{\mathcal E}{{\mathcal E}_c}\right)^2\,
\exp \left[ -\pi\, \frac{{\mathcal E}_c}{\mathcal E} \right]\,,
\end{equation} 
apart from an ``inessential'' (c.\,f. Ref.~\cite{Brezin:1970})  
pre-exponential factor of $\pi$. 
In this case, $e^+e^-$ pair production at a laser has all the features of 
the usual tunneling effect.
On the other hand, for $\gamma\gg 1$, i.\,e. in the low-field, high-frequency
limit, formula~(\ref{brezin-rate}) ressembles a perturbative result, 
\begin{equation}
 w_{\rm B\,I} 
\simeq 
\frac{c}{4\,\pi^3 \lambdabar^4} 
\left( \frac{\hbar\,\omega}{m_e c^2}\right)^2
\frac{\pi}{2\gamma\ln(4\gamma )}
\left(
\frac{\rm e}{4\gamma}
\right)^{2\,\frac{2\,m_e\,c^2}{\hbar\,\omega}}
\left( 1+\mathcal O \left(\frac{1}{\gamma^2}\right)\right)\,,
\hspace{4ex} {\rm \ for\ } \gamma\gg 1\,.
\label{brezin-rate-weak}
\end{equation}
Since $\gamma\propto 1/e$, formula~(\ref{brezin-rate-weak}) corresponds to the 
$n$-th order perturbation theory, $n$ being the minimum number of quanta
required to create an $e^+e^-$ pair: $n\gwig 
2\, m_e c^2/(\hbar\omega)\gg 1$. Therefore, expression~(\ref{brezin-rate}) for 
the pair production rate interpolates 
analytically between the adiabatic, nonperturbative tunneling mechanism 
($\gamma\ll 1$) and the anti-adiabatic, perturbative multi-photon production 
mechnism ($\gamma\gg 1$).

The principal, exponential factor, 
$\exp [-\pi({\mathcal E}_c/\mathcal E)g(\gamma )]$, in the pair production 
probability~(\ref{brezin-rate}) has been confirmed by later 
work~\cite{Popov:1971,Popov:1972,Popov:1974,Popov:1974b}. 
However, the imaginary-time method, exploited by these later studies, allowed
to determine the pre-exponential factor more accurately, by taking into 
account also interference effects. 
It was found~\cite{Popov:1974,Popov:1974b} that the pair production 
probability $w$, under the conditions~(\ref{conditions}), can be represented as
a sum of probabilities $w_n$ of many-photon processes, 
\begin{equation}
\label{popov-rate}
w_{\rm P}  = \sum_{n>n_0} w_n\,, 
\hspace{6ex} {\rm with\ }\  n_0 =\frac{m_e c^2}{\hbar \omega}\,\triangle\,,
\end{equation}
where the latter are given\footnote{For notationally simplicity, we use 
$\hbar = c =1$ in Eq.~(\ref{w_n}).} in terms of an integral over the 
three momentum components 
$\mathbf p = ({\mathbf p}_\perp,p_\parallel)$, of the produced $e^-$ 
(or, equivalently, $e^+$), perpendicular (${\mathbf p}_\perp$)  and 
parallel ($p_\parallel$) to the applied electric 
field~(\ref{electric-type}), respectively,   
\begin{eqnarray}
\label{w_n}
w_n &= & \frac{2}{\pi}\, \omega^2\, 
\exp\left[
-\pi \frac{{\mathcal E}_c}{\mathcal E} 
g(\gamma )
\right]
\int \frac{{\rm d}^3p}{(2\pi )^3}
\left[ 
1 - (-)^n  
\cos\left( 4\,\frac{p_\parallel}{\omega}\,  \arctan \gamma \right) 
\right] 
\times
\\[1ex] \nonumber
&& \times\,
\exp\left[
-\pi \frac{{\mathcal E}_c}{\mathcal E}
\left\{ 
\left( g(\gamma ) + \frac{1}{2} \gamma g^\prime (\gamma )\right)
 \frac{{\mathbf p}_\perp^2}{m_e^2} 
- \gamma\frac{\rm d}{{\rm d}\gamma} 
\left(  g(\gamma ) + \frac{1}{2} \gamma g^\prime (\gamma ) \right) 
\frac{p_\parallel^2}{m_e^2}
\right\}
\right]
\times
\\[1ex] \nonumber
&& \times\ 
\delta\left( 
\triangle (\gamma )  + 
\frac{1}{2}\left( 1 +\gamma \frac{\rm d}{{\rm d}\gamma}
\right) \triangle (\gamma )\, \frac{{\mathbf p}_\perp^2}{m_e^2} 
+\frac{1}{2}
\left( 1 -\gamma \frac{\rm d}{{\rm d}\gamma}
\right)
\left( 1 +\gamma \frac{\rm d}{{\rm d}\gamma}
\right) \triangle (\gamma )\,  
\frac{p_\parallel^2}{m_e^2} -n\,\frac{\omega}{m_e} \right)
\,.
\end{eqnarray}
In Eqs.~(\ref{popov-rate}) and (\ref{w_n}), the function 
$\triangle (\gamma )$,
\begin{equation}
\triangle (\gamma ) = \frac{4}{\pi} 
\frac{\sqrt{1+\gamma^2}}{\gamma} \,
\mbox{\boldmath $E$} 
\left( \frac{1}{\sqrt{1+\gamma^2}}\right)
=
\left\{
\begin{array}{lcr}
\frac{4}{\pi\gamma}\left( 1 -\frac{1}{2}\gamma^2 \ln \left(
\frac{\gamma}{4\sqrt{\rm e}}\right) +\mathcal O(\gamma^4)\right)
&:& \gamma\ll 1\,,\\[2ex]
2+\frac{1}{2}\frac{1}{\gamma^2}+\mathcal O(1/\gamma^4) 
&:& \gamma\gg 1\,,
\end{array}
\right.
\end{equation}
plays the r\^ole of an effective gap width between the lower and upper
continuum (in units of $m_e c^2$). The presence of the factor 
$\cos\left( 4\,(p_\parallel/\omega)\, \arctan \gamma \right)$
in the rate~(\ref{w_n}) results in oscillations\footnote{Analogous oscillations
play an important r\^ole in the description of (p)reheating after
inflation~\cite{Kofman:1997yn}.} in the momentum spectrum, 
mostly manifest at $p_\parallel=0$ (see also 
Refs.~\cite{Narozhnyi:1974,Mostepanenko:1974}). In this case 
$w_n\propto 1-(-)^n$, i.\,e. electrons are created by odd harmonics. This 
selection rule is due to interference and related to particle statistics. 
The argument of the of the 
delta function in formula~(\ref{popov-rate}) expresses conservation
of energy and determines the dispersion law for the electron in the 
strong external field. 

The integration over the three-momentum $\mathbf p$ in formula~(\ref{w_n})
can be performed analytically and the result can be expressed in terms of 
special functions~\cite{Popov:1974b}. For our purpose, however, we need only 
the limiting cases of small and large $\gamma$, which are even simpler,
\begin{eqnarray}
\label{w_popov_lim}
\lefteqn{
w_{\rm P} \simeq  \frac{c}{4\,\pi^3 \lambdabar^4}\,\times}
\\[1ex] \nonumber && 
\times\,\left\{
\begin{array}{lcr}
\frac{\sqrt{2}}{\pi} 
 \left( 
\frac{\mathcal E}{{\mathcal E}_c}
\right)^{\frac{5}{2}}\,
\exp \left[ 
-\pi\, \frac{{\mathcal E}_c}{\mathcal E} 
\left( 1-\frac{1}{8}\gamma^2+\mathcal O(\gamma^4)\right)
\right]\,,
&:& \gamma\ll 1\,,\\[2ex]
\sqrt{\frac{\pi}{2}} 
\left( 
\frac{\hbar\,\omega}{m_e c^2}
\right)^{\frac{5}{2}}
\sum_{n>2\frac{m_ec^2}{\hbar \omega}}
\left( \frac{{\rm e}}{4\gamma}\right)^{2n}
{\rm e}^{ 
-2\left( n-2\frac{m_ec^2}{\hbar \omega}\right)}
{\rm Erfi} \left( 
\sqrt{2\left( n-2\frac{m_ec^2}{\hbar \omega}\right)}\right)
&:& \gamma\gg 1\,,
\end{array}
\right.
\end{eqnarray}
where $\rm Erfi$ is the imaginary error function~\cite{Erdelyi}. Both
asymptotics bear a neat similarity to the corresponding asymptotics
of formula~(\ref{brezin-rate}). Moreover, the result~(\ref{w_popov_lim}) is, 
in the adiabatic limit ($\gamma\ll 1$), in agreement with the classical
result~(\ref{schwinger-rate}) of Schwinger, if one properly averages
the latter over an oscillation period.

The conditions~(\ref{conditions}) give an indication on the range of 
validity of the quoted results. In 
Refs.~\cite{Popov:1974,Popov:1974b,Marinov:1977gq} it has been argued that 
for $\mathcal E\,\gwig\, 0.1\,{\mathcal E}_c$ the backreaction of the 
produced $e^+e^-$ pairs on the external field and the mutual interactions
between these particles has to be taken into account 
These effects in the superstrong field regime, which are expected to lead to 
the formation of a plasma, can in principle be accounted for 
with the help of methods developed in 
Refs.~\cite{Novikov:1980ni,Spokoinyi:1982pg,Cooper:1989kf,Kluger:1991ib,
Kluger:1992gb,Kluger:1993md,Kluger:1998bm,Schmidt:1998vi,Schmidt:1999zh}, 
mainly in the context of particle production in the central rapidity region in
heavy ion collisions. The discussion of the plasma regime is however beyond 
the scope of the present letter, in which we just want to estimate the 
onset of the Schwinger mechanism. 

%%%%%%%%%%%%%%%%%%%%%%%%%%%%%%%%FIGURE%%%%%%%%%%%%%%%%%%%%%%%%%%
\begin{figure}
\begin{center}
\parbox{13.9cm}{\includegraphics*[width=13.9cm]{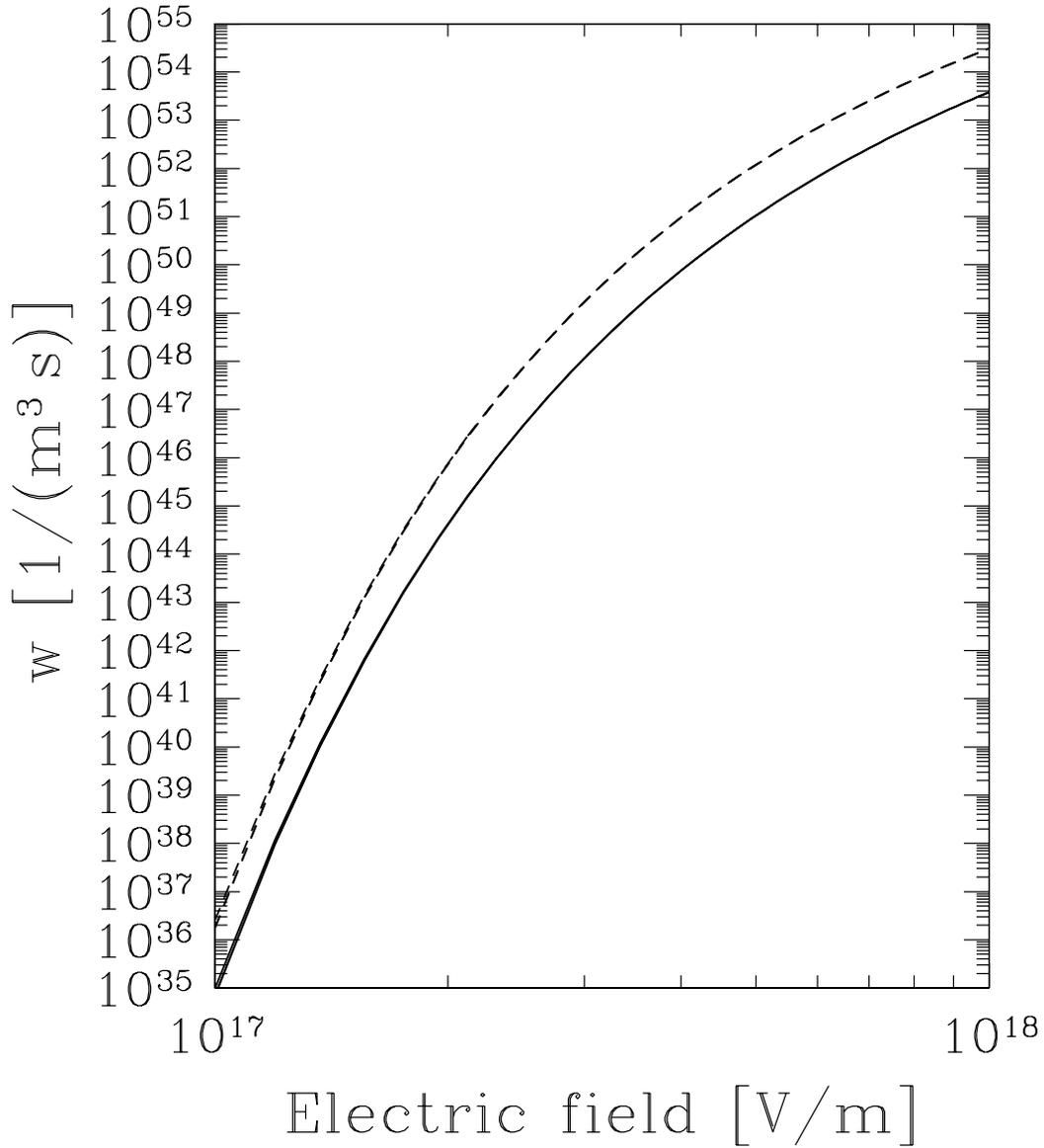}}
\caption[dum]{\label{rate-fig} 
The probability $w$ to produce an $e^+e^-$ pair   
per unit volume and unit time at a laser, 
with $E_\gamma =\hbar\,\omega\,\leq\,10$ keV, as a function of the peak
electric field $\mathcal E$. The solid and dashed lines refer to the 
estimates~(\ref{popov-rate}) and (\ref{brezin-rate}), respectively. 
The dependence on the laser frequency 
$\omega$ is very weak and only visible at the lower end of the considered
range of the electric field. 
The respective lower curves correspond to $E_\gamma =1$ keV, the upper
ones to $E_\gamma =10$ keV. We note, that the probability that an $e^+e^-$ 
pair is produced within one Compton space-time volume is very small,  
$w\cdot\lambdabar^4/c \approx w\cdot 10^{-58}\ {\rm m^3\, s}\ll 1$, in the 
whole range of $\mathcal E$ considered. 
}
\end{center}
\end{figure}
%%%%%%%%%%%%%%%%%%%%%%%%%%%%%%%%%%%%%%%%%%%%%%%%%%%%%%%%%%%%%%%%%

%\section{Rate Estimates: Phenomenology}
{\em 4.}
Let us apply now the theoretical rate estimates reviewed above in order to 
determine the critical laser parameters which should be aimed at to get an 
observable effect. 

In Fig.~\ref{rate-fig} we display the probability density $w$ 
as a function of the peak electric field $\mathcal E$, for 
laser photon energies $\hbar\,\omega\,\leq\,10$ keV, 
as applicable at presently planned lasers (c.\,f. Table~\ref{parameters}).  
In order to appreciate the scale of $w$ in Fig.~\ref{rate-fig}, we note
that the Compton space-time volume of an electron has the size
\begin{equation}
\label{compton_sp_t_vol}
\triangle V_e\times \triangle T_e = 
\lambdabar^3\times (\lambdabar/c)\ \simeq\  
7.4\times 10^{-59}\ {\rm m}^3\,{\rm s}\,.
\end{equation}
In other words, the probability that an $e^+e^-$ pair is produced within one 
Compton space-time volume is very small in Fig.~\ref{rate-fig}, 
$w\,\triangle V_e \,\triangle T_e\ll 1$, as it should be under the 
conditions~(\ref{conditions}). 
Furthermore, we observe that formula~(\ref{brezin-rate}) overestimates the 
rate in the considered range of $\mathcal E$ by about one order of
magnitude. Last but not least we note that   
in the considered range of the electric field, $\mathcal E\gwig 10^{17}$ V/m, 
the dependence on the laser frequency $\omega$ is 
very weak and nearly invisible. This is due to the fact that 
the adiabatic, nonperturbative, strong field regime, $\gamma\,\lwig\, 1$, 
starts to apply for 
$\mathcal E\,\gwig\, \mathcal E_\omega \equiv \hbar\omega\, 
{\mathcal E}_c/(m_ec^2)\sim 10^{15\div 16}$ V/m 
(c.\,f. Eq.~(\ref{gamma})), for $\hbar\,\omega\sim 1\div 10$~keV.

\begin{figure}
\begin{center}
\parbox{13.9cm}{\includegraphics*[width=13.9cm]{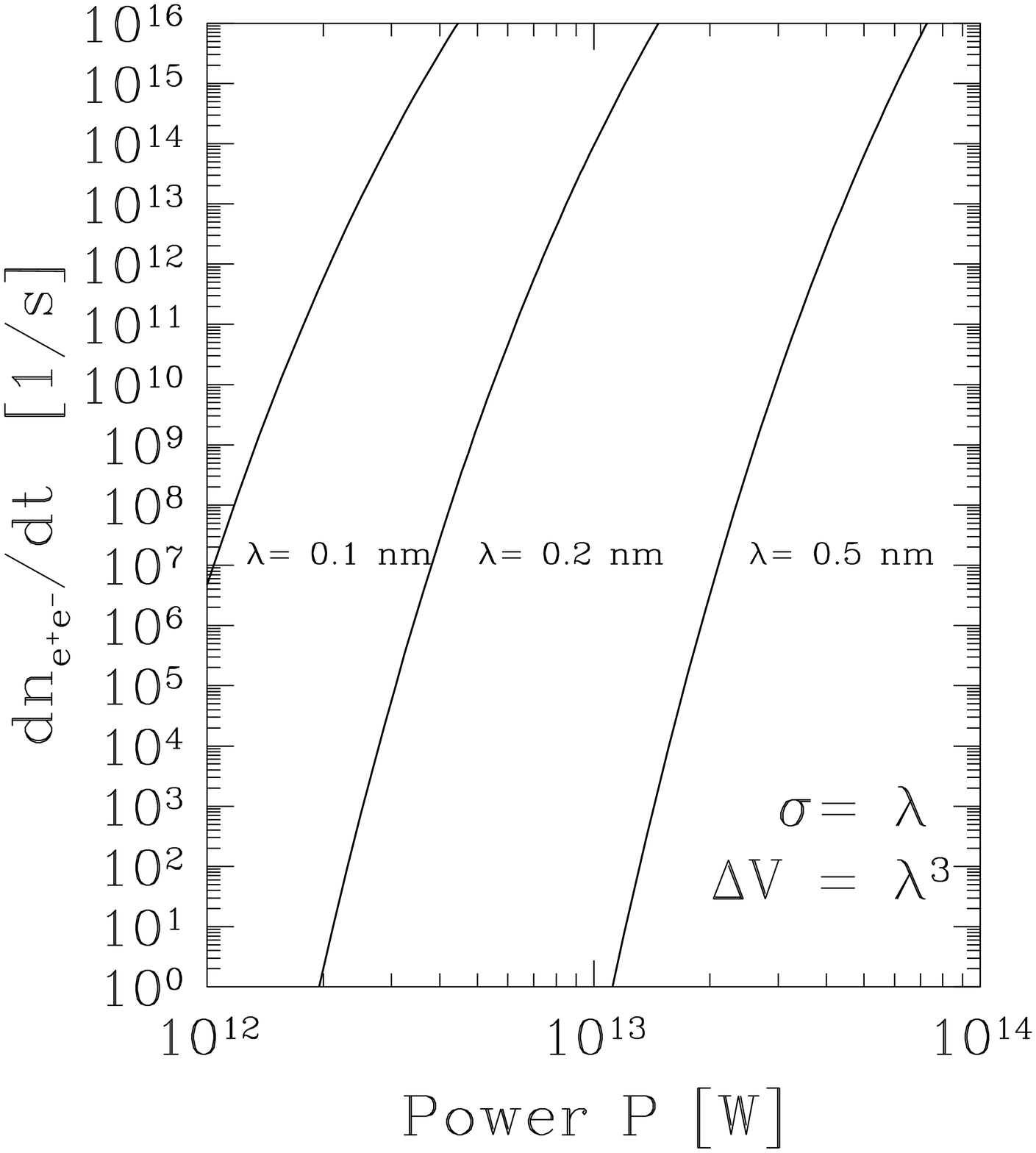}}
\caption[dum]{\label{n-rate-fig} 
The average number~(\ref{n-rate}) of $e^+e^-$ pairs produced per unit time,
$w\,\triangle V$, 
where formula~(\ref{popov-rate}) is taken for the pair production probability 
$w$, as a function of the laser power $P$ for different values of the laser 
wavelength $\lambda$. Here it is assumed that the diffraction limit, 
$\sigma = \lambda$ and $\triangle V = \lambda^3$, can be reached. 
}
\end{center}
\end{figure}

Let $\triangle V =\sigma^3$ be the effective volume, where the peak electric 
field~(\ref{peak-electric-field}) exists, and consider the average number of 
$e^+e^-$ pairs per unit time, produced in this volume,
\begin{equation}
\label{n-rate}
\frac{{\rm d}\,n_{e^+e^-}}{dt} 
= w\,\triangle V\,.
\end{equation}
In Fig.~\ref{n-rate-fig} we plot this quantity as a function of the laser
power $P$ for different laser spot radii $\sigma$, under the assumption
that the diffraction limit of focusing, $\sigma = \lambda$ and 
$\triangle V = \lambda^3$, can be reached. From this figure and the 
typical duration times $\triangle t\sim 10^{-(13\div 16)}$ s of the
coherent laser pulses presently discussed (c.\,f. Table~\ref{parameters}), we 
infer that a minimum power of $2.5\div 4.5$ TW, corresponding to a  
electric field of $(1.7\div 2.3)\times 10^{17}$ V/m, is needed to produce at 
least one $e^+e^-$ pair, $n_{e^+e^-}=w\,\triangle V\triangle t\,\sim\, 1$, 
if the laser has a wavelength of 0.1 nm and the theoretical diffraction limit 
is actually reached. For larger wavelength or in the case that the 
diffraction limit cannot be reached ($\sigma \gg \lambda$), as is presently
the case with existing technology~\cite{Graeff:priv,Hastings:2000} 
(c.\,f. column labeled ``Focus: Available'' in Table~\ref{parameters}), 
this critical power increases considerably. For example, for a spot radius   
$\sigma = 20$ nm, the critical power rises to $38\div 55$~PW, corresponding 
to a electric field of $(1.1\div 1.3)\times 10^{17}$ V/m.
%%%%%%%%%%%%%%%%%%%%%%%%%%%%%%%%FIGURE%%%%%%%%%%%%%%%%%%%%%%%%%%
%%%%%%%%%%%%%%%%%%%%%%%%%%%%%%%%%%%%%%%%%%%%%%%%%%%%%%%%%%%%%%%%%

The minimally required power $P_{\rm min}$, the corresponding  power 
density $S_{\rm min}$ and electric field ${\mathcal E}_{\rm min}$, to produce
at least one $e^+e^-$ pair in a volume $\triangle V = \sigma^3$ during 
a time interval $\triangle t$ at the focus of a laser with spot radius 
$\sigma$ are given in Table~\ref{crit-parameters}. This table demonstrates 
clearly that in order to get a sizeable effect it is mandatory to 
reach the diffraction limit of focusing at X-ray lasers. We note that 
research and development in this direction is under 
way~\cite{tesla-tdr,Hastings:2000}.
%%%%%%%%%%%%%%%%%%%%%%%%%%%Table%%%%%%%%%%%%%%%%%%%%%%%%%%%%%%%%%%%%%%
\begin{table} 
{\footnotesize
\begin{center}
\begin{tabular}{|l|c|c|c||c|c|c|}\hline
& $\lambda$ &  $\sigma $
 & $\triangle t$  & $P_{\rm min}$ & $S_{\rm min}$
& ${\mathcal E}_{\rm min}$
\\\hline\hline  
Focused X-ray FEL: & 0.1 nm 
&  0.1 nm& 0.1 ps &  2.5 TW & $7.8\times 10^{31}$ W/m$^2$ 
& $1.7\times 10^{17}$ V/m\\
($\approx$ ``Goal'' in Table~\ref{parameters})   
&0.1 nm & 0.1 nm& 0.1 fs &  4.5 TW& $1.4\times 10^{32}$ W/m$^2$ 
&$2.3\times 10^{17}$ V/m\\\hline
Focused X-ray FEL: & 0.1 nm & 20 nm &0.1 ps  &  38 PW
& $3.0\times 10^{31}$ W/m$^2$ 
&$1.1\times 10^{17}$ V/m\\
($\approx$ ``Available'' in Table~\ref{parameters})
& 0.1 nm & 20 nm& 0.1 fs &  55 PW& $4.3\times 10^{31}$ W/m$^2$ 
&$1.3\times 10^{17}$ V/m\\\hline\hline
Focused Optical Laser: &1 $\mu$m & 
1 $\mu$m & 10 ps & 49 EW & $1.6\times 10^{31}$ W/m$^2$ 
&$7.7\times 10^{16}$ V/m\\
Diffraction Limit &1 $\mu$m &
1 $\mu$m & 100 fs & 58  EW & $1.8\times 10^{31}$ W/m$^2$ 
&$8.3\times 10^{16}$ V/m\\\hline
\end{tabular}
\caption[dum]{This table displays the minimally required power $P_{\rm min}$, 
the corresponding power density $S_{\rm min}$ and electric field 
${\mathcal E}_{\rm min}$, to produce
at least one $e^+e^-$ pair in a volume $\triangle V = \sigma^3$ during 
a time interval $\triangle t$ at the focus of a laser with spot radius 
$\sigma$ (1 EW = 10$^{18}$ W). 
\label{crit-parameters}} 
\end{center}
}
\end{table}
%%%%%%%%%%%%%%%%%%%%%%%%%%%%%%%%%%%%%%%%%%%%%%%%%%%%%%%%%%%%%%%%%%%%%%

As discussed at the end of the last section, for 
$\mathcal E > {\mathcal E}_{\rm min}$ the system is expected to enter rather
rapidly the plasma regime. In this sense, the power
densities $S_{\rm min}$ in Table~\ref{crit-parameters} represent
also theoretical upper limits for laser fields. 

%\section{Discussion and Outlook}
{\em 5.}
We conclude that 
the power densities and electric fields which can be reached with presently
available technique (column labeled ``Focus: Available'' in 
Table~\ref{parameters}) are far too small to lead to a sizeable effect. On 
the other hand, if X-ray optics can be considerably improved, allowing the 
theoretical diffraction limit to be reached, and if the FEL power can be 
increased from the presently planned values to the 
terawatt regime, as has been argued to be possible by the tapered undulator 
technique~\cite{Chen:1998}, then there is 
ample room (c.\,f. column labeled ``Focus: Goal'' in Table~\ref{parameters}) 
for an investigation of the Schwinger pair production mechanism at a future
X-ray FEL. 
Intensive development in technical areas, particularly in that of X-ray
optics, will be needed in order to achieve the required ultrahigh 
power densities~\cite{tesla-tdr,Hastings:2000}.
It should be pointed out, however, that even though progress to achieve such 
a lofty goal is rather slow and laborious, the rewards that may be derived in 
this unique regime are so extraordinary that looking into XFEL's or 
LCLS's extension to this regime merits serious considerations.
There will be unprecedented opportunities to use these intense X-rays in order
to explore some issues of fundamental physics that have eluded
man's probing so far~\cite{Tajima:2000}.
 
\section*{Acknowledgements}
I wish to thank P. Zerwas for drawing my attention to this subject.
Furthermore, many discussions with, and fruitful comments from, my
colleagues at HASYLAB, W. Graeff and G. Materlik, are acknowledged.
Moreover, I thank P. Chen, C. Pellegrini and T. Tajima for encouragement,
and J. Baacke, W. Buchm\"uller, M.~G.~Schmidt and F.~Schrempp for a careful 
reading of the manuscript.

\end{document}